\documentclass[aps,prl,onecolumn,showpacs,groupedaddress]{revtex4}
\usepackage{graphicx}

\begin{document}

\title{Generating entangled coherent state of two cavity modes in three-level $\Lambda$-type atomic system}
\author{Qing-Xia Mu, Yong-Hong Ma, L.Zhou }
\affiliation{ \\School of physics and optoelectronic technology,
Dalian University of Technology, Dalian 116024 China
                     }

\date{\today}

\begin{abstract}
In this paper, we present a scheme to generate an entangled coherent
state by considering a three-level $"\Lambda "$ type atom
interacting with a two-mode cavity driven by classical fields. The
two-mode entangled coherent state can be obtained under large
detuning condition. Considering the cavity decay, an analytical
solution is deduced.
\end{abstract}

 \maketitle
\subsection{I. Introduction}
Entanglement between quantum systems is recognized nowadays as a key
ingredient for testing quantum mechanics versus local hidden-variable theory %
\cite{Hybrid-32}. Entanglement as a valuable resource has been used
in quantum information processing such as quantum computation
\cite{Hybrid-1}, quantum sweeping and teleportation \cite{Hybrid-2}.
As macroscopic nonclassical states, Schr$\ddot{o}$dinger cat states
and entangled coherent states  have always been an attractive topic.
In quantum optics, these two kinds of states are described as
superpositions of different coherent states and superpositions of
two-mode coherent states, respectively. It has been shown that such
superposition states have many practical applications in quantum
information processing \cite{Hybrid-4}. So far, a variety of
physical systems presenting entangled
coherent states have been investigated  \cite%
{Hybrid-7,Hybrid-24,Hybrid-9,Hybrid-10,Hybrid-25,Hybrid-29,Hybrid-12}.
Sanders \cite{Hybrid-7} presented a method for generating an
entangled coherent state with equal weighting factors by using a
nonlinear Kerr medium placed in one arm of the nonlinear
Mach-Zehnder interferometer.  Wielinga \emph{et al.}
\cite{Hybrid-24} modified this scheme via an optical tunnelling
device instead of the Kerr medium to generate entangled coherent
states with a variable weighting factor. Schemes have also been
proposed for generating such entangled coherent states using trapped
ions \cite{Hybrid-9} by controlling the quantized ion motion
precisely.
\begin{figure}[tbp]
\includegraphics*[width=70mm,height=90mm]{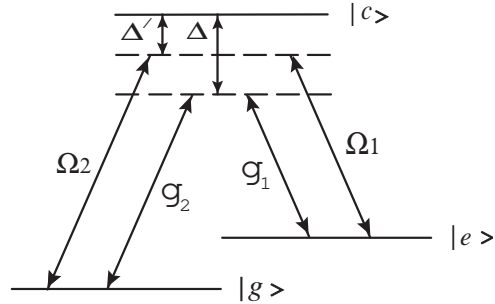}
\caption{Schematic diagram of a three-level $\Lambda -$type atom
interacting
with two cavity modes and two classic fields with detunings $\Delta $ and $%
\Delta ^{\prime }$, respectively.}
\end{figure}

On the other hand, cavity QED, with Rydberg atoms interacting with
an electromagnetic field inside a cavity, has also been proved to be
a promising environment to generate quantum states. In the context
of cavity QED, several schemes have been proposed to generate such
superposition coherent states \cite
{Hybrid-10,Hybrid-25,Hybrid-29,Hybrid-12}. Ref. \cite {Hybrid-25}
showed that entangled coherent states can be generated by the
state-selective measurement on a two-level atom interacting with a
two-mode field. Recently, Wang and Duan \cite{Hybrid-29} studied the
generation of multipartite and multidimensional cat states by
reflecting coherent pulses successively from a single-atom cavity.
Solano \emph{et al.} \cite{Hybrid-12} proposed a method for
generating entangled coherent states by considering a two-level atom
cavity QED driven by a strong classic field. However, the two cavity
modes in this scheme interact with the same atomic transitions, and
thus can not be easily manipulated.

In our research, we present an alternative method to prepare two
modes of cavity in an entangled coherent state with the context of
cavity QED. Based on the nonresonant interaction of a three-level
$"\Lambda"$ type atom with two cavity modes and two classic fields,
we can obtain the entangled coherent states. Compared with Ref.
\cite{Hybrid-12}, the two cavity modes in our research interact with
different atomic transitions so that they are easy to be recognized
and manipulated. Furthermore, we work on the large detuning
condition, so the decoherence induced by the spontaneous emission of
excited level $|c\rangle $ can be ignored. Our scheme can also be
generalized to generate multidimensional entangled coherent state
with the assistance of another two-level atom in two-photon process.

\subsection{II. The theoretical model and calculation}

The system we consider is a three-level atom in $\Lambda $
configuration placed inside a two-mode field cavity. The level
structure of the atom is depicted in Fig.1, where the two atomic
transitions $|c\rangle \leftrightarrow |e\rangle $ and $|c\rangle
\leftrightarrow |g\rangle $ interact with the two cavity modes with
the same detuning $\Delta $ but with different coupling constants
$g_{1}$ and $g_{2}$, respectively. The two atomic transitions
$|c\rangle \leftrightarrow |e\rangle $ and $|c\rangle
\leftrightarrow |g\rangle $ are also driven by two classical fields
with detuning $\Delta ^{\prime }$, and $\Omega _{1}$ and $\Omega
_{2}$ are the Rabi frequencies of the two classical fields. The
Hamiltonian for the system can be written as
\begin{eqnarray}
H &=&\hbar w_{e}|e\rangle \langle e|+\hbar w_{c}|c\rangle \langle
c|+\hbar
w_{1}a_{1}^{\dagger }a_{1}+\hbar w_{2}a_{2}^{\dagger }a_{2}  \nonumber \\
&&+\hbar g_{1}(a_{1}^{\dagger }|e\rangle \langle c|+a_{1}|c\rangle
\langle e|)+\hbar g_{2}(a_{2}^{\dagger }|g\rangle \langle
c|+a_{2}|c\rangle \langle
g|)  \nonumber \\
&&+\hbar \Omega _{1}(e^{-i(w_{c}-w_{e}-\Delta ^{\prime })t}|c\rangle
\langle e|+H.c.)+\hbar \Omega _{2}(e^{-i(w_{c}-\Delta ^{\prime
})t}|c\rangle \langle g|+H.c.),
\end{eqnarray}
where $a_{i}^{\dagger }$ and $a_{i}$ are the creation and
annihilation operators for the cavity fields of frequencies
 $w_{i}$ (i=1,2), while $w_{c}$ and $w_{e}$ are the Bohr frequencies
associated with the two atomic transitions $|c\rangle
\leftrightarrow |g\rangle $ and $|e\rangle \leftrightarrow |g\rangle
$, respectively.

We consider the large detuning domain
\begin{eqnarray}
\left( \frac{\Omega _{1}}{\Delta^{\prime } },\frac{\Omega _{2}}{%
\Delta ^{\prime }},\frac{g_{1}}{\Delta }, \frac{g_{2}}{\Delta
}\right) \ll 1.
\end{eqnarray}
After adiabatically eliminating the excited level $|c\rangle $, we
derive the effective Hamiltonian as follows \cite{Hybrid-14}
\begin{eqnarray}
H_{eff}{=}-\hbar g_{eff}(a_{1}^{\dagger }a_{2}\sigma ^{\dagger
}+a_{1}a_{2}^{\dagger }\sigma )-\hbar \Omega _{eff}(\sigma ^{\dagger
}+\sigma ),
\end{eqnarray}
where $g_{eff}{=}\frac{g_{1}g_{2}}{\Delta }$, $\Omega
_{eff}{=}\frac{\Omega _{1}\Omega _{2}}{\Delta ^{\prime }}$; $\sigma
^{\dagger }{=}\left| e\right\rangle \left\langle g\right| $ and
$\sigma {=}\left| g\right\rangle \left\langle e\right| $ are raising
and lowering atomic operators, respectively. In Eq.(3) we have
assumed that the Stark shifts can be corrected by retuning the laser
frequencies \cite{Hybrid-22}.

In the strong driving regime $\Omega _{eff}{\gg }g_{eff}$, we choose $%
H_{eff}^{0}{=}-\hbar \Omega _{eff}(\sigma ^{\dagger }+\sigma )$ and $%
H_{eff}^{I}{=}-\hbar g_{eff}(a_{1}^{\dagger }a_{2}\sigma
^{+}+a_{1}a_{2}^{\dagger }\sigma )$. By performing the unitary
transformation $U{=}e^{-\frac{i}{\hbar }H_{eff}^{0}t}$ on
$H_{eff}^{I}$, in which we neglect the terms that oscillate with
high frequencies, the Hamiltonian reads
\begin{eqnarray}
H_{eff}^{int}=-\frac{\hbar g_{eff}}{2}(a_{1}^{\dagger
}a_{2}+a_{1}a_{2}^{\dagger })(\sigma ^{\dagger }+\sigma ).
\end{eqnarray}
We recognize the field Hamiltonian part $-\frac{\hbar g_{eff}}{2}%
(a_{1}^{\dagger }a_{2}+a_{1}a_{2}^{\dagger })$ is the generator of
the SU(2) coherent state \cite{Hybrid-27}. Here, we are interested
in using the Hamiltonian of Eq.(4) to entangle the two cavity modes
through the interaction with the atom. For this purpose we consider
the case that the atom state is initially prepared in the ground
state $|g\rangle $, while
both of the two cavity fields are in coherent states $|\alpha \rangle $ and $%
\beta \rangle $, respectively. Thus the initial state of the system
is
\begin{eqnarray}
|\Psi (0)\rangle =|g\rangle \otimes |\alpha ,\beta \rangle .
\end{eqnarray}
 On the basis of $|\pm \rangle
=\frac{1}{\sqrt{2}}(|g\rangle \pm|e\rangle )$,
which are the eigenstates of $\sigma +\sigma ^{\dagger }$ with eigenvalues $%
\pm 1$, the time evolution of the system is given by
\begin{eqnarray}
&|\Psi (t)\rangle &=e^{\frac{-i}{\hbar }H_{eff}^{int}t}|\Psi
(0)\rangle \nonumber
\\
&&=\frac{1}{\sqrt{2}}e^{\frac{ig_{eff}t}{2}(K_{+}+K_{-})}|+,\alpha
,\beta \rangle
+\frac{1}{\sqrt{2}}e^{\frac{-ig_{eff}t}{2}(K_{+}+K_{-})}|-,\alpha
,\beta \rangle ,
\end{eqnarray}%
where $K_{+}=a_{1}^{\dagger }a_{2}$, $K_{-}=a_{1}a_{2}^{\dagger }$.
These
operators satisfy the SU(2) commutation relations, i.e. $%
[K_{-},K_{+}]=-2K_{0}$, $[K_{0},K_{+}]=K_{+}$, $[K_{0},K_{-}]=-K_{-}$, with $%
K_{0}=\frac{1}{2}(a_{1}^{\dagger }a_{1}-a_{2}^{\dagger }a_{2})$.
Thus we can use the SU(2) Lie algebra \cite{Hybrid-15} to expand the
unitary evolution operator $e^{\pm
\frac{ig_{eff}t}{2}(K_{+}+K_{-})}$ as
\begin{equation}
e^{\pm \frac{ig_{eff}t}{2}(K_{+}+K_{-})}=e^{\pm x_{+}K_{+}}e^{K_{0}\ln {x_{0}%
}}e^{\pm x_{-}K_{-}},
\end{equation}%
in which
\begin{eqnarray*}
x_{0} &=&\{\cosh {\frac{ig_{eff}t}{2}}\}^{-2}, \\
x_{+} &=&x_{-}=\tanh {\frac{ig_{eff}t}{2}}.
\end{eqnarray*}%
Using Eq.(7) we can conveniently derive the evolution of the system
as
\begin{equation}
|\Psi (t)\rangle =\frac{1}{\sqrt{2}}|+\rangle |\tilde{\alpha},\tilde{\beta}%
\rangle +\frac{1}{\sqrt{2}}|-\rangle |\tilde{\alpha}^{\ast },\tilde{\beta}%
^{\ast }\rangle ,
\end{equation}%
with%
\begin{eqnarray*}
\tilde{\alpha} &=&\alpha \cos {\frac{g_{eff}t}{2}}+i\beta \sin {\frac{%
g_{eff}t}{2}}, \\
\tilde{\beta} &=&\beta \cos {\frac{g_{eff}t}{2}}+i\alpha \sin {\frac{g_{eff}t%
}{2}}.
\end{eqnarray*}
We now change the basis back to original atomic states
\begin{eqnarray}
|\Psi (t)\rangle {=}\frac{1}{2}|g\rangle (|\tilde{\alpha},\tilde{\beta}%
\rangle +|\tilde{\alpha}^{\ast },\tilde{\beta}^{\ast }\rangle )+\frac{1}{2}%
|e\rangle (|\tilde{\alpha},\tilde{\beta}\rangle -|\tilde{\alpha}^{\ast },%
\tilde{\beta}^{\ast }\rangle ). \end{eqnarray} When the atom comes
out from the two-mode cavity, we can use level-selective ionizing
counters to detect the atomic state. If the internal state of atom
is detected to be in the state $|g\rangle $ or $|e\rangle $, Eq.(9)
will project the two-mode cavity into
\begin{equation}
|\Psi _{f}(t)\rangle {=}\frac{1}{\sqrt{M}}(|\tilde{\alpha},\tilde{\beta}%
\rangle \pm |\tilde{\alpha}^{\ast },\tilde{\beta}^{\ast }\rangle ),
\end{equation}%
where $M$ is normalization factor such that
\begin{eqnarray}
M=2\pm \lbrack exp(-|\tilde{\alpha}|^{2}-|\tilde{\beta}|^{2}+\tilde{\alpha}%
^{\ast ^{2}}+\tilde{\beta}^{\ast ^{2}})+exp(-|\tilde{\alpha}|^{2}-|\tilde{%
\beta}|^{2}+\tilde{\alpha}^{2}+\tilde{\beta}^{2})].
\end{eqnarray}
By this way we obtain a superposition of two two-mode coherent
states. It is interesting to note that under certain conditions on
the amplitudes of two coherent states, such superposition state can
exhibit nonclassical effects such as violation of the
Cauchy-Schwartz inequality and two-mode squeezing \cite{Hybrid-16}.
On the other hand, the interaction time of the atom in the cavity
can be controlled as $m\pi /g_{eff}$ by using a velocity selector,
where $m$ is odd number. Then we can obtain two-mode even and odd
coherent states as $|\Psi _{f}(t)\rangle
{=}\frac{1}{\sqrt{M}}(|i\beta ,i\alpha \rangle \pm |-i\beta
,-i\alpha \rangle )$ \cite{Hybrid-16}. It has been proved that these
even and odd coherent states exist strong correlations between two
modes.

Now we try to estimate the entanglement of Eq.(10). Recently,
different
entanglement criteria for two-mode systems have been proposed in \cite%
{Hybrid-20,Hybrid-18,Hybrid-19}. Here, we choose constructing
normalized and orthogonal basis and then use concurrence to evaluate
the entanglement proposed in \cite{Hybrid-20,Hybrid-26}. According
to Ref. \cite{Hybrid-26}, the concurrence of Eq.(10) is given by
\begin{eqnarray}
C=\frac{2}{|M|}\sqrt{(1-|p_{1}|^{2})(1-|p_{2}|^{2})}.
\end{eqnarray}
where $P_{1}=e^{-|\tilde{\alpha}|^{2}+\tilde{\alpha}^{\ast ^{2}}}$ and $%
P_{2}=e^{-|\tilde{\beta}|^{2}+\tilde{\beta}^{\ast ^{2}}}$.
\begin{figure}[tbp]
\includegraphics*[width=70mm,height=70mm]{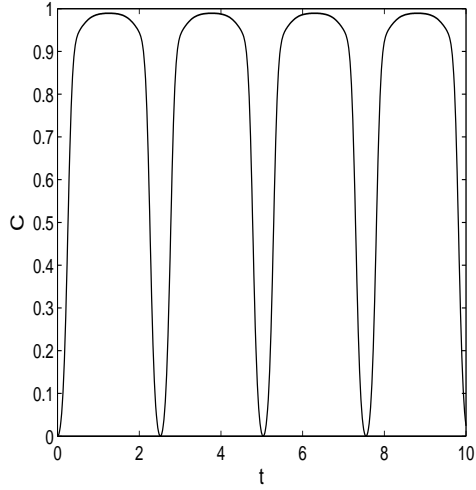}
\caption{The time evolution of the degree of the entanglement with $%
g_{eff}=2.5$, $\protect\alpha =1$, $\protect\beta =1.5$.}
\end{figure}

Fig.2 shows the time evolution of the concurrence. Here the positive
sign has been chosen for Eq.(10). We see that under this group of
parameters of the two modes, concurrence oscillates periodically
with time. From Eq.(10), it
is easy to see that the state is entangled at any other time, except when $%
\tilde{\alpha}$ and $\tilde{\beta}$ are real, namely $t=n\pi
/g_{eff}$ (where $n$ is even number).

\subsection{III. Analytical solution including cavity decay}

Due to the large detuning, the excited atomic level $|c\rangle $ do
not participate in the interaction. Therefore, the spontaneous
emission atomic level can be ignored. Now, we discuss the time
evolution of the system under the cavity losses. For simplicity, we
assume the losses of the two cavity modes are equal. By including
the cavity damping terms in the equation of motion for the density
operators, the mast equation can be written as
\begin{eqnarray}
\dot{\rho}=\frac{-i}{\hbar }[H_{eff},\rho ]+L_{1}\rho +L_{2}\rho ,
\end{eqnarray}
where $L_{i}=\frac{k}{2}(2a_{i}\rho a_{i}^{\dagger }-a_{i}^{\dagger
}a_{i}\rho -\rho a_{i}^{\dagger }a_{i})$ for $i=1,2$.

This equation can be solved by Lie algebras \cite{Hybrid-15} and
superoperator technique \cite{Hybrid-23}. When the initial state is
prepared in $|g,\alpha,\beta\rangle$, we can obtain the analytical
solution of the system as follows
\begin{eqnarray}
\rho&=&\frac{1}{2}|+,\tilde{\alpha}e^{\frac{-kt}{2}},\tilde{\beta}e^{\frac{%
-kt}{2}}\rangle \langle +,\tilde{\alpha}e^{\frac{-kt}{2}},\tilde{\beta}e^{%
\frac{-kt}{2}}|+\frac{1}{2}|-,\tilde{\alpha}^*e^{\frac{-kt}{2}},\tilde{\beta}%
^*e^{\frac{-kt}{2}}\rangle \langle-,\tilde{\alpha}^*e^{\frac{-kt}{2}},\tilde{%
\beta}^*e^{\frac{-kt}{2}}|  \nonumber \\
&&+\frac{1}{2}\eta |+,\tilde{\alpha}e^{\frac{-kt}{2}},\tilde{\beta}e^{\frac{%
-kt}{2}}\rangle \langle -,\tilde{\alpha}^*e^{\frac{-kt}{2}},\tilde{\beta}%
^*e^{\frac{-kt}{2}}|+\frac{1}{2}\eta^* |-,\tilde{\alpha}^*e^{\frac{-kt}{2}},%
\tilde{\beta}^*e^{\frac{-kt}{2}}\rangle\langle +,\tilde{\alpha}e^{\frac{-kt}{%
2}},\tilde{\beta}e^{\frac{-kt}{2}}|,  \nonumber \\
\end{eqnarray}
where
\begin{eqnarray}
\eta{=}exp[-4\lambda_1\tilde{\alpha}\tilde{\beta}+(|\tilde{\alpha}|^2 +|%
\tilde{\beta}|^2)(e^{-kt}-1)+2\lambda_2(\tilde{\alpha}^2+\tilde{\beta}^2)],
\nonumber
\end{eqnarray}
\begin{eqnarray}
\lambda_1{=}\frac{kg_{eff}\cos(g_{eff}t)-k^2\sin(g_{eff}t)-kg_{eff}e^{-kt}}{%
2i(k^2+g_{eff}^2)},  \nonumber
\end{eqnarray}
\begin{eqnarray}
\lambda_2{=}\frac{k^2\cos(g_{eff}t)+kg_{eff}\sin(g_{eff}t)-k^2e^{-kt}}{%
2(k^2+g_{eff}^2)}.
\end{eqnarray}
Then we measure the atomic state in the bare basis $\{|g\rangle,|e\rangle\}$%
. If the atom is detected in the ground state $|g\rangle$, the field
will be projected into the state
\begin{eqnarray}
\rho_f&=&\frac{1}{N}[|\tilde{\alpha}e^{\frac{-kt}{2}},\tilde{\beta}e^{\frac{%
-kt}{2}}\rangle \langle \tilde{\alpha}e^{\frac{-kt}{2}},\tilde{\beta}e^{%
\frac{-kt}{2}}|  \nonumber \\
&&+\eta |\tilde{\alpha}e^{\frac{-kt}{2}},\tilde{\beta}e^{\frac{-kt}{2}%
}\rangle \langle \tilde{\alpha}^*e^{\frac{-kt}{2}},\tilde{\beta}^*e^{\frac{%
-kt}{2}}|  \nonumber \\
&&+\eta^*|\tilde{\alpha}^*e^{\frac{-kt}{2}},\tilde{\beta}^*e^{\frac{-kt}{2}%
}\rangle\langle \tilde{\alpha}e^{\frac{-kt}{2}},\tilde{\beta}e^{\frac{-kt}{2}%
}|  \nonumber \\
&&+|\tilde{\alpha}^*e^{\frac{-kt}{2}},\tilde{\beta}^*e^{\frac{-kt}{2}%
}\rangle \langle\tilde{\alpha}^*e^{\frac{-kt}{2}},\tilde{\beta}^*e^{\frac{-kt%
}{2}}|],
\end{eqnarray}
where $N$ is the normalization coefficient
\begin{eqnarray}
N=2+\eta \exp[(-|\tilde{\alpha}|^2-|\tilde{\beta} |^2
+\tilde{\alpha}^2+\tilde{\beta}
^2)e^{-kt}]+\eta^*\exp[(-|\tilde{\alpha}|^2-|\tilde{\beta}|^2+
\tilde{\alpha}^{*^2}+\tilde{\beta}^{*^2})e^{-kt}].
\end{eqnarray}

The time dependent factors $\eta $ and $\eta ^{\ast }$ are more
important and interesting here. They contain the information how
fast the density matrix becomes an incoherent mixture state. Then we
still use concurrence to estimate the entanglement. The normalized
and orthogonal basis is defined as
\begin{eqnarray*}
\text{For cavity mode }1,|0\rangle &{=}&|\tilde{\alpha}e^{\frac{-kt}{2}%
}\rangle ,|1\rangle {=}\frac{|\tilde{\alpha}^{\ast
}e^{\frac{-kt}{2}}\rangle
-p_{1}|\tilde{\alpha}e^{\frac{-kt}{2}}\rangle }{M_{1}}, \\
\text{For cavity mode }2,|0\rangle &{=}&|\tilde{\beta}e^{\frac{-kt}{2}%
}\rangle ,|1\rangle {=}\frac{|\tilde{\beta}^{\ast
}e^{\frac{-kt}{2}}\rangle
-p_{2}|\tilde{\beta}e^{\frac{-kt}{2}}\rangle }{M_{2}}.
\end{eqnarray*}%
with $p_{1}{=}\exp [(-|\tilde{\alpha}|^{2}+\tilde{\alpha}^{\ast
^{2}})e^{-kt}]$, $M_{1}=\sqrt{1-|p_{1}|^{2}}$, $p_{2}{=}\exp [(-|\tilde{\beta%
}|^{2}+\tilde{\beta}^{\ast ^{2}})e^{-kt}]$,
$M_{2}=\sqrt{1-|p_{2}|^{2}}$.

After calculation, the entanglement of system $\rho _{f}$ has the
form
\begin{eqnarray}
C=\frac{2M_{1}M_{2}}{N}|\eta |.
\end{eqnarray}
\begin{figure}[tbp]
\includegraphics*[width=80mm,height=80mm]{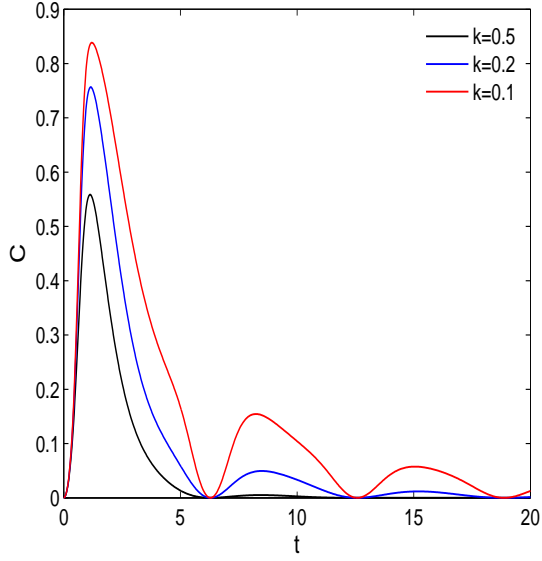}
\caption{The time evolution of the entanglement when considering
cavity decay with $g_{eff}$=1, $\protect\alpha =1$, $\protect\beta
=1.5$. From top to bottom, $k=0.1,0.2,0.5$, respectively.}
\end{figure}
Fig.3 displays the entanglement of two cavity modes measured by
concurrence for $k=0.1,0.2,0.5$, respectively. It is observed that
amplitude of concurrence decreases with the increasing of $k$. The
loss of the cavity destroys the entanglement. Thus, a high-Q
two-mode cavity is preferred.

Furthermore, our method can also be extended to generate
multidimensional entangled coherent state. In order to do this, we
first send a two-level atom with a virtual intermediate level
\cite{Hybrid-30}, initially in the ground state $|g\rangle $,
through a two-mode cavity. The atom dispersively interact with one
of the cavity mode(e.g.,cavity mode  with annihilation (creation)
operators $a_{1}(a_{1}^{\dagger }))$ where the two-photon process
takes place. The effective Hamiltonian acting on state $|g\rangle $ is $%
H=-\hbar \lambda a_{1}^{\dagger }a_{1}(a_{1}^{\dagger }a_{1}-1)$ \cite%
{Hybrid-31}. If the cavity mode is initially in a coherent state,
the
nonlinear Hamiltonian interaction equals to that of the Kerr medium \cite%
{Hybrid-28}. When the two-level atom flies out of the cavity, a
three-level atom in $\Lambda $ configuration is sent into it. Doing
the same operation we discussed in section II, finally we recognize
that the total evolution
operator of the field part has the same form as Eq.(4) in Ref. \cite%
{Hybrid-28}. Following the methods of Ref. \cite{Hybrid-28}, we can
derive the multidimensional entangled coherent state after a
projective measurement of atomic state in the basis $\{|\pm \rangle
\}$.

\subsection{IV. Conclusion}

In conclusion, we present a scheme to generate two-mode entangled
coherent state via the QED system, in which a three-level $"\Lambda
"$ configuration atom interacts with two cavity modes and two
classic fields in large detuning. When we perform a measurement on
the atomic state, the two-mode field will collapse into the
entangled coherent state if the two cavity modes are both in the
coherent states initially. In our scheme the two cavity modes
interact with two distinct atomic transitions, so they are easy to
be controlled. Moreover, taking into account the cavity decay, we
study the system evolution and give an analytical solution. With the
assistance of another two-level atom with intermediate level, our
scheme can also be generalized to generate multidimensional
entangled coherent state.

\end{document}